# The Case for a Collaborative Universal Peer-to-Peer Botnet Investigation Framework

**Mark Scanlon and Tahar Kechadi**
**UCD School of Computer Science and Informatics, University College Dublin, Belfield, Dublin 4, Ireland**
mark.scanlon@ucd.ie
tahar.kechadi@ucd.ie

**Abstract:** Peer-to-Peer (P2P) botnets are becoming widely used as a low-overhead, efficient, self-maintaining, distributed alternative to the traditional client/server model across a broad range of cyberattacks. These cyberattacks can take the form of distributed denial of service attacks, authentication cracking, spamming, cyberwarfare or malware distribution targeting on financial systems. These attacks can also cross over into the physical world attacking critical infrastructure causing its disruption or destruction (power, communications, water, etc.). P2P technology lends itself well to being exploited for such malicious purposes due to the minimal setup, running and maintenance costs involved in executing a globally orchestrated attack, alongside the perceived additional layer of anonymity. In the ever-evolving space of botnet technology, reducing the time lag between discovering a newly developed or updated botnet system and gaining the ability to mitigate against it is paramount. Often, numerous investigative bodies duplicate their efforts in creating bespoke tools to combat particular threats. This paper outlines a framework capable of fast tracking the investigative process through collaboration between key stakeholders.

## 1. Introduction

Discussions and reporting of technology savvy criminals attacking financial institutions, governments and corporate assets have become regular news stories in recent years. Many of these attacks target specific institutions motivated by political or financial goals. These cyberattacks can be aided by a number of different advances in technology in recent years but one of the most threatening technologies to be exploited for these malicious purposes is the capability to execute crimes remotely using botnets. Botnets take the form of massive globally distributed networks of zombie slave machines capable of executing the commands of their creator, the botmaster. In financial terms, these distributed networks are largely responsible for the executing of numerous financially motivated crimes, such as online payment account and bank account cracking, spamming, distributed denial of service attacks, etc. Riccardi et al. (2012) found that the Zeus botnet alone is estimated to have caused damages of over $100 million USD since its discovery in 2007 through the targeting of financial websites.

Of course the dangers of distributed cyberattacks are not solely limited to financial crimes. Increasingly botnets are becoming the weapon of choice in new online battlefields. Parks and Duggan (2011) defined cyberwarfare as cyberattacks with kinetic or non-cyber world affects. Cyberwarfare is a very real component of modern conflict. Cyberwarfare attacks are commonly occurring between countries that are not physically at ware. For example, cyberwarfare attacks from North Korea on the U.S. were found to be up 800% in the period from 2005-2007 by the U.S. Department of Homeland Security (Lee, 2011).

The Anonymous "hacktivist" group attacks in recent years have displayed the power of distributed computational contribution to botnet powered attacks. The contribution to these attacks is not solely innocent victims' machines haven been taken control of by a botmaster - regular Internet users with shared political or activist views voluntarily decided to contribute their processing power to a collaborative cause. Partaking in the Anonymous attacks involved users downloading and configuring an open source network stress-testing tool called "Low Orbit Ion Cannon" (LOIC). Alternatively, users had the option to donate their computational power via a JavaScript-based version facilitating anyone who visits the site to participate in the attack (Mansfield-Devine, 2011). Regular P2P file-sharing networks, such as BitTorrent, can also be manipulated by malicious users to aid in the execution of a DDoS attack through the exploitation of vulnerabilities in the protocol and operation of the network (El Defrawy et al., 2007). In a cyberwarfare scenario, it is conceivable that citizens of countries with limited computational infrastructure or supporters of terrorist organizations could similarly be called upon to donate their systems to aid in a collaborated attack on an enemy's infrastructure. Throughout the world, a significant amount of effort is taken in an attempt to combat malicious botnet activity. Much of this effort is duplicated with different investigative and research entities each creating bespoke tools focused on a particular botnet system, or on a particular botnet variant. This paper proposes a botnet investigation framework capable of eliminating much of this duplicated effort and improving the overall reaction time of the botnet forensic community.

### 1.1 Motivations behind cyberattacks

The potential for attacks originating from a growing number of sources is a concern for the security of many nations across the globe. Amoroso (2012) defined five possible motivations behind cyberattacks:

- Country-sponsored warfare - This is whereby enemy cyber forces attack national infrastructure in an attempt to disable critical resources of the opposing country in a similar manner to traditional physical warfare. Only the resources and devotion of the attacking nation limit the intensity of this attack. In a P2P botnet facilitated attack, citizens could voluntarily donate their computing power towards the national goals in a similar manner to the Anonymous attacks outlined above.

- Terrorist Attack - Groups driven by terrorist motivations could quickly gain sufficient funding and expertise to conduct their attacks. Also in this scenario, regular Internet users could partake in a terrorist operation without requiring any skill or expertise by donating their regular computer equipment to the attack.

- Commercially motivated attack - Competing companies might target their competitors' e-commerce infrastructure in order to prevent regular users from purchasing anything from their online stores. A popular e-commerce site being taken offline has the added effect of harming the victim company's reputation in their customers' eyes.

- Financially driven criminal attack - These types of attacks could target individual computer users by recording their Internet banking details, online payment services and other financial services. Companies can also be targeted with extortion threats against their online infrastructure.

- Hacking - This scenario generally involves an individual or group of hackers attacking targets motivated by little more than mischief or attaining online recognition of their achievements.

### 1.2 P2P networks

Since P2P networking has become mainstream, the technology has been deployed across a broad range of systems and services. While the level of variation in topologies is significant, all P2P networks must share a number of common attributes:

- Capability to connect to the network (bootstrapping) - When a new node wishes to join the network, it must have the ability to contact at least one other active participant in the network. Depending on the network design, this may take the form of a hardcoded list of active nodes (typical in a decentralized topology) or a list of bootstrapping servers (typical in a centralized topology) (Conrad & Hof, 2007).

- Record Maintenance of Active Nodes - In a decentralized network the peers themselves must all contribute to the recording of active nodes on the network. No single peer has the entire list, with each peer contributing to a collective distributed database, typically a distributed hash table (DHT). In a centralized design, this duty falls on the controlling server(s). As each new node comes online, it announces its presence to the database maintainer and requests a list of other active peers to begin working.

- Query/Order/File Propagation - In order for a P2P network to fulfill whatever the purpose it was designed for, intra-peer communication is requisite. As a result of this necessity, each peer must be able to receive requests or commands and pass these communications onto other known peers.

- Software Maintenance - The P2P enabled binary can quickly become outdated. The upgrade process must be simple to perform while maintaining node uptime. While newer versions of the application might have additional functionality, it must to ensure backwards compatibility otherwise the network as a whole may suffer.

In this paper, we introduce a framework that enables forensic investigators and researchers to fast track the investigation of any P2P network. The framework exploits many of the common attributes of these networks outlined above. Each node on a P2P network has two main functions. The first involves participating in the maintenance of the network itself -- it is aware of a number of other nodes, and is in communication with a subset of the overall population. From the analysis of the P2P communication of an active node in the network, common communications can be identified, e.g., peer discovery, query/command/file propagation, etc. Once each of these communication patterns is identified, the behavior of an active node can be recorded in a centralized shared database. The sharing of identified patterns will aid in the elimination of duplicated work by forensic investigators. As the contribution to the framework increases, both the database of identifiable traffic patterns and the value of using the framework will become significantly greater.

## 2. Premise for universal investigation framework

The collaborative investigative framework proposed can take advantage of the fundamental commonality in design of most P2P botnet systems:

- The fundamental requirement for intra-peer communication resulting in any compromised machine gaining the ability to communicate with other peers on the network.
- The ability to self-propagate to further vulnerable machines.
- The requirement for each newly infected machine to connect with the botmaster or to the botnet itself to receive its latest order.
- The requirement for each active node in the network to maintain a local list of other active nodes.
- The compulsory command forwarding responsibility of each node.

Wang et al. (2009) defined the characteristics of common botnet designs into three main types of P2P botnets: parasitic, leeching and bot-only. Parasitic botnets take advantage of vulnerable hosts within an existing P2P network with the upper limit of the botnet's growth being the number of vulnerable machines on the network. Leeching botnets refers to designs that depend on an existing P2P network for their communication. A full list of the characteristics of 3 botnet variations can be seen in Table 1 below.

**Table 1:** Comparison of three types of P2P Botnets

| Features | Parasite | Leeching | Bot-only |
|---|---|---|---|
| Infection Vectors | P2P malware | Any kind of malware | Any kind of malware |
| Bot Candidates | Vulnerable hosts in P2P networks | Vulnerable hosts in the Internet | Vulnerable hosts in the Internet |
| Bootstrap Procedure | None | Required | Required |
| Members in the Network | Legitimate peers and bots | Legitimate peers and bots | Only bots |
| Communication Protocols | Existing P2P protocols | Existing P2P protocols | Self-designed/existing P2P protocols |
| Command & Control Styles | Pull or Push | Pull or Push | Pull or Push |

The framework should be designed in such a modular manner to facilitate the convenient plug in of the characteristics and communication signatures of any particular P2P botnet. Standard investigation types, such as botnet enumeration, node classification, botmaster eavesdropping and potential botmaster identification, can all be included as part of the framework. The conduction of any of the aforementioned investigations would require the appropriate protocol and communication signature being recorded for each of the differing networks. The framework itself could take control of the investigation from that point. The key to the success of such a framework lies in a sufficiently comprehensive centralized database of botnets being contributed to by interested stakeholders.

Traditionally, the time required from the identification of a botnet to gaining the ability to partake in the network might require hundreds, if not thousands, of man-hours. It is envisioned that using a standardized framework can greatly expedite the process. With sufficient contribution from network security and forensic experts in the identification and characterizing of botnet signatures, the framework outlined in this paper will have the ability to quickly facilitate the investigation of any known botnet and to quickly adapt to the discovery of new networks.

Another significant advantage to such a framework is to aid in the elimination of the duplication of efforts. Botnet researchers around the world from law enforcement, public institutions and private corporations would each benefit from the collaborative system. P2P botnets aid criminals in the distributed perpetration of their crimes - it seems fair that the targets and protectors against these crimes should also have a unified global network to combat these crimes.

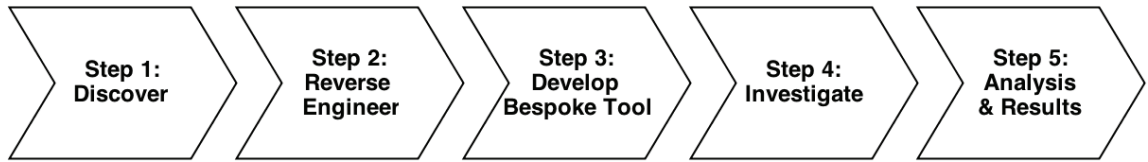

**Figure 1:** Steps Involved in a typical P2P Botnet investigation

## 3. Universal framework system components

The framework is designed to be as modular as possible. This facilitates its expandability in the future as more networks and components are added. The core components of the system are outlined below:

- Network Signatures - The signatures for a given P2P botnet will have two components; a specification of the communication commands available to any node on the network and a parameter file. The specification will store the formats required for each command required for intra-peer chatter. The parameter file will define the network and its features, e.g., DHT, peer exchange, communication frequencies, etc.

- Client Emulator - This module takes the above network signatures and connects to the botnet in question. The methods and frequency of communication with each peer will be provided from the signature. The purpose of this component is to ensure the appearance of the framework on the network. To each other node, it is imperative that any investigative tool appears as though it is any other regular node on the network.

- Investigation Controllers - Manipulation of the parameters of the emulator can result in the execution of the differing investigation types outlined above. Each investigation type will have its own controller module that is responsible for the investigation.

- Evidence Storage - The evidence collected will be stored in an evidence bag. Both the network stream captured during the investigation alongside the interpreted evidence gathered by partaking on the network. The integrity of this evidence is paramount and is described in greater detail in the Section 4.5.

## 4. P2P network investigation types

Once the communication method of the undocumented network is reverse engineered through traditional means and the required signatures stored in the database, crawling the network will then be possible by any contributing agency or institution. The framework is designed to aid in the following investigation types:

### 4.1 Evidence collection

In order to combat the unauthorized downloading of copyrighted material, many countries have implemented a three to six strikes ``graduated response'' system whereby repeat offenders will have their Internet service discontinued for a defined penalty period (Serbin, 2012). In order for such systems to operate, evidence must be gathered to prove that infringement has taken place. In a botnet investigation, the evidence collected might take the form of the commands issued, the origin of these commands, the targets of an attack, etc. The concept of evidence collection is easily understood with respect to illegal file sharing and botnet investigations, but it can equally apply to more legitimate P2P applications such as Voice over IP (VoIP) services, e.g., Skype (Baset & Schulzrinne, 2004), or instant messaging services, e.g., AIM or MSN Messenger (Jennings et al., 2006).

### 4.2 Anatomy

Investigating the anatomy of a particular network involved the analysis of the client binary's behaviour and analysis of the network communication patterns. This type of investigation attempts to classify the system as centralized/decentralized, Client-Server/P2P hybrid or solely P2P based command and control. An anatomy documentation investigation can continue past the network architecture of the system to cover some of the counter-detection and anti-forensic techniques employed. For example, Goel et al. (2006) discovered that ``Agobot'' had a built in defence mechanism to terminate the execution of a remotely upgradable list of over 610 anti-virus programs. The Storm worm was also engineered to aggressively use the distributed botnet to collectively attack anyone who attempted to reverse engineer it (Mukamurenzi, 2008).

### 4.3 Wide-area measurement

This concentrates on attempting to enumerate the population of the network, its bandwidth, its computational power or its often-multifaceted goals. Gathering the population of a P2P network is traditionally a non-trivial task as the number of nodes ever connecting to any single node or to a Command and Control (C&C) server may only count for a small proportion of the participating nodes. Rajab et al. (2007) specified that there are two definitions of a P2P network's size:

- Footprint - This indicates the aggregated total number of machines that have been compromised over time.

- Live Population - This measure denotes the number of compromised machines that are concurrently in communication with each other.

In the case of measuring a documented network, e.g., BitTorrent, a custom crawler must be built which can efficiently collect peer information. In the case of an undocumented network (e.g. a P2P botnet), a relatively straightforward method for measuring the population of the network is to run a bot on a deliberately infected machine and monitor the resultant network traffic. The number of IP addresses the infected node is in communication with can be easily counted having eliminated all non-botnet related network traffic. It is unsafe to assume that a single node will ultimately communicate with every other node in the network over any reasonable time frame. Traditional enumeration investigation increase the number of infected machines (physically or virtually) and amalgamate the results should lead to a more accurate representation. Using the framework, regular client usage can be amplified to achieve the required results in a significantly timelier manner.

### 4.4 Takeover

Botnet takeover involves a third party gaining control of a botnet from its owner. This third party could be law enforcement, researchers or another cybercriminal. Once control has been gotten of the botnet, the new botmaster is able to issue commands, update configurations and operate the botnet as desired. Stone-Gross et al. (2009) successfully took over the Torpig botnet for 10 days. During this time, the researchers identified more than 180,000 compromised machines and were sent over 70GB of automatically harvested personal information. The number of discovered unique Torpig bot IDs and corresponding number of IP addresses was observed to be 182,914 and 1,247,642 respectively (Stone-Gross et al., 2011). The discrepancy between the number of bots and IP addresses found is accountable by network effects such as Dynamic Host Configuration Protocol (DHCP) churn and Network Address Translation (NAT), as described in greater details in Section 5 below.

### 4.5 Forensic integrity

Due to the sensitive nature of digital evidence collection, it is imperative that the data collected by any forensic tool is absolutely verifiable and identical to the original source. This integrity is ensured in the framework through the implementation of a new live digital evidence bag designed to operate with P2P network evidence collection. This evidence bag will record all pertinent information, e.g., IP addresses, packets, running processes, etc., alongside the raw network traffic capture. By partaking in the P2P network itself, the evidence gathered does not need to be reverse engineered, i.e., all of the information available to a regular client of that network will be amplified and recorded in this evidence bag. Once any network traffic is collected, each packet is time-stamped and logged. The time stamping facilitates real-time event reconstruction packet by packet, emulating the original traffic.

The integrity of any evidence collected by the framework will be insured through the implementation of regular hash checking on the data being collected using SHA-512 (Secure Hashing Algorithm producing a 512-bit long hash). The system collects a stream of information, the stream itself is hashed and both are stored on the external drive or can be uploaded to secure cloud storage. During the transmission process, the integrity of each of the chunks being transferred is maintained due to a SHA-512 hash being computed as the chunk is being transmitted. Server-side, once the transmission is completed, a SHA-512 hash is taken on the chunk and verified against the original. If these hashes do not match, i.e., the integrity of that chunk has been compromised in transmission, a failure notification is sent to the client, which queues that chunk up again for transmission.

*Mark Scanlon and Tahar Kechadi*## 5. Advantages

Implementing a universal collaborative framework quickly and easily facilitates different investigative bodies conducting similar investigations worldwide. Some of the specific advantages are outlined below:

- Compatibility - Having a centralized framework facilitates the integration with numerous other network forensic tools as required. Expandability is also ensured through the modularization of the components and specification of the individual networks.

- Cost - The cost savings for participating in such a collaborative framework are significant. Eliminating the duplicated development of bespoke tools will greatly reduce the resources required.

- Automated Identification - With a sufficiently large database of known networks established automated identification.

- Cross-border cooperation - Differing interests across national borders should be advantageous tot he system as a whole. Differing bodies developing the signatures required for their own operational requirements will result in a more complete robust database of signatures.

- Speed - The regular P2P botnet investigation process can be seen in Figure 1. Using the framework described, the entire investigation process is fast-tracked as far as Step 4 for each contributing body (assuming the network is already documented). The final analysis phase can also be expedited through common evidence processing procedures.

## 6. Potential issues

As with any system designed with a requirement for collaboration between differing global bodies and organizations, there are a number of potential issues:

- Key Stakeholder Buy-in - Without a sufficient contribution towards the framework (both the core functionality and the botnet signatures), the system can rapidly become out-dated in comparison with the quick progression in botnet technologies.

- Access Maintenance and Control - A framework that potentially has the ability to infiltrate and take control over any reverse engineered botnet is something that would be highly valuable if it got into the wrong hands. Controlling access to such a framework is paramount to its reliability and usefulness to law enforcement and forensic investigators. One solution to this issue would be for the framework to require server-side authentication/update before any investigation can commence.

- Corporate Participation - While having collaboration from private corporations would be advantageous to the system as a whole, from a corporate perspective having exclusivity on the solution to any problem is fundamental to many business models. As a result, it is envisioned that the system will primarily be contributed to and used by law enforcement and research institutions.

## 7. Conclusions and future work

A proof-of-concept framework has been created capable of integrating with a number of common P2P networks, such as BitTorrent, with modules implemented for each of their constituent components. This proof-of-concept needs to be expanded to cover a broader range of technologies, such as modules for the popular distributed hash tables. The ever-growing range of crimes perpetrated by global distributed networks of zombie machines is staggering. Having key stakeholders such as researchers, digital investigators, law enforcement and security experts pool and share their expertise and efforts can significantly aid in the on-going battle against cyberattacks.

## Acknowledgements

This work has been co-funded by the Irish Research Council and Intel Ireland Ltd. through the Enterprise Partnership Scheme. The authors also wish to acknowledge the contribution of Amazon for supporting this research through an Amazon Web Services research grant.

## References

Amoroso, E. G. 2012, Cyber attacks: protecting national infrastructure, Elsevier.
Baset, S. A. & Schulzrinne, H. 2004, 'An analysis of the skype peer-to-peer internet telephony protocol'.

Conrad, M. & Hof, H.-J. 2007, 'A generic, self-organizing, and distributed bootstrap service for peer-to-peer networks', , 59-72.
El Defrawy, K., Gjoka, M., & Markopoulou, A. 2007, BotTorrent: misusing BitTorrent to launch DDoS attacks, USENIX Association.
Goel, S., Baykal, A., & Pon, D. 2006, 'Botnets: the anatomy of a case', Journal of Information Systems Security.
Jennings, R. B., Nahum, E. M., Olshefski, D. P., Saha, D., Shae, Z.-Y., & Waters, C. 2006, 'A study of internet instant messaging and chat protocols', Network, IEEE, 20, 4, 16-21.
Lee, S. 2011, 'US Security Strategy toward North Korea's Cyber Terrorism', Carnegie Institution for Science.
Mansfield-Devine, S. 2011, 'Anonymous: serious threat or mere annoyance?', Network Security, 2011, 1, 4-10.
Mukamurenzi, N. M. 2008, 'Storm Worm: A P2P Botnet'.
Parks, R. C. & Duggan, D. P. 2011, 'Principles of Cyberwarfare', Security & Privacy, IEEE, 9, 5, 30-35.
Rajab, M. A., Zarfoss, J., Monrose, F., & Terzis, A. 2007, My botnet is bigger than yours (maybe, better than yours): why size estimates remain challenging, USENIX Association.
Riccardi, M., Pietro, R. D., Palanques, M., & Vila, J. A. 2012, 'Titans' revenge: detecting Zeus via its own flaws', Computer Networks.
Serbin, D. 2012, 'The Graduated Response: Digital Guillotine or a Reasonable Plan for Combating Online Piracy?', Intellectual Property Brief, 3, 3, 4.
Stone-Gross, B., Cova, M., Cavallaro, L., Gilbert, B., Szydlowski, M., Kemmerer, R., Kruegel, C., & Vigna, G. 2009, Your botnet is my botnet: analysis of a botnet takeover, ACM.
Stone-Gross, B., Cova, M., Gilbert, B., Kemmerer, R., Kruegel, C., & Vigna, G. 2011, 'Analysis of a botnet takeover', Security & Privacy, IEEE, 9, 1, 64-72.
Wang, P., Wu, L., Aslam, B., & Zou, C. C. 2009, A systematic study on peer-to-peer botnets, IEEE.